\documentclass[]{spie}  
\usepackage[dvips]{graphicx}

\title{Seeing Statistics at the Upgraded 3.8m UK Infrared Telescope (UKIRT)} 

\author{Marc S. Seigar\supit{1}, Andy J. Adamson\supit{1}, 
Nicholas P. Rees\supit{1}, Timothy G. Hawarden\supit{2}, \\
Malcolm J. Currie\supit{3}, Timothy C. Chuter\supit{1}
\skiplinehalf
\supit{1}Joint Astronomy Centre, 660 N. A'ohoku Place, Hilo, HI 96720, USA 
\skiplinehalf
\supit{2}Astronomy Technology Centre, Royal Observatory, Blackford Hill, 
Edinburgh, EH9 3HJ, UK
\skiplinehalf
\supit{3}Starlink, Rutherford Appleton Laboratory, Chilton, Didcot, Oxon,
OX11 0QX, UK
}

\authorinfo{Further author information: (Send correspondence to M.S.S.)\\M.S.S.: E-mail: m.seigar@jach.hawaii.edu, Telephone: (808) 969-6565}

\begin{document} 
  \maketitle 

\begin{abstract}
From 1991 until 1997, the 3.8m UK Infrared Telescope (UKIRT) underwent a
programme of upgrades aimed at improving its intrinsic optical
performance. This resulted in images with a FWHM of
0.\hspace*{-1mm}$^{\prime\prime}$17 at 2.2 $\mu$m in September 1998.
To understand and maintain the improvements to the delivered
image quality since the completion of the upgrades programme, we have
regularly monitored the overall {\it atmospheric} seeing, as measured by
radial displacements of subaperture images (i.e. seeing--generated focus
fluctuations), and the {\it delivered} image diameters. The latter have
been measured and recorded automatically since the beginning of 2001
whenever the facility imager UFTI (UKIRT Fast Track Imager) has been in
use.

In this paper we report the results of these measurements. We investigate
the relation between the delivered image diameter and the RMS atmospheric
seeing (as measured by focus fluctuations, mentioned above). 
We find that the best seeing occurs in the second half of the night, generally
after 2am HST and that the best seeing occurs in the summer between the 
months of July and September. We also find that the relationship between
$Z_{rms}$ and delivered image diameter is uncertain. As a result $Z_{rms}$
frequently predicts a larger FWHM than that measured in the images.

Finally, we show that there is no correlation between near--infrared seeing 
measured at UKIRT and sub--mm seeing measured at the Caltech Submillimetre 
Observatory (CSO).
\end{abstract}


\keywords{Telescope optical quality, facility seeing, sub--mm seeing}

\section{INTRODUCTION}
\label{sect:intro}  

Between 1991 and 1997 the UK Infrared Telescope (UKIRT) was the subject of a 
systematic campaign of improvements, the UKIRT upgrades programme, which had
the explicit goal of providing imaging performance competitive with the best
ground--based facilities (Hawarden et al. 1994, 1996, 1998). By active control
of the primary--mirror figure and secondary--mirror alignment (Chrysostomou et
al. 1998) and tip--tilt actuation of the secondary mirror under control of a
fast guider, this goal has been met (Hawarden et al. 1999, 2000). The telescope
routinely delivers images with FWHM less than 
0.\hspace*{-1mm}$^{\prime\prime}$6.
In 1998 a programme to measure the delivered seeing was undertaken. Images
were taken with the Infrared Camera, IRCAM, and the median seeing was
found to be 0.\hspace*{-1mm}$^{\prime\prime}$45.
Some images obtained in this programme in September 1998, 
had a delivered FWHM of 0.\hspace*{-1mm}$^{\prime\prime}$17.

The new facility imager, the UKIRT Fast Track Imager (UFTI), was delivered at
the end of 1998. At the beginning of 2001, a programme to regularly monitor
the delivered seeing with UFTI was initiated. In this paper we present results
from this programme and compare them with pure atmospheric seeing,
as measured from the RMS in the focus position, $Z_{rms}$.

\section{IMAGE QUALITY}
\label{sec:dsee}

Throughout 2001, UKIRT's delivered image quality was regularly monitored with
UFTI.  Whenever UFTI is in use the {\tt ORAC-DR} imaging pipeline (Economou
et al. 1999, Currie 2001) automatically performs photometry on all
standard--star frames taken, and a Gaussian fit is used to
estimate the FWHM of the central source.

The pipeline calculates the best--fitting elliptical Gaussian
point--spread function (PSF) using the Starlink {\tt KAPPA} 
(Currie \& Berry 2000)
package. The {\tt PSF} task forms marginal profiles at
45$^{\circ}$--intervals about the weighted centroid (from task {\tt
CENTROID}) within a 9.\hspace*{-1mm}$^{\prime\prime}$2 square.  
It rejects outliers from any
neighbouring sources by ensuring that the intensity declines radially,
and excludes bad pixels.  For each profile, PSF fits a Gaussian curve,
iteratively solving normal equations after background (lower quartile)
subtraction.  The combined profiles yield approximate values for the
Gaussian width, axis ratio and orientation.  {\tt PSF} averages the pixel
data into isophotal elliptical annuli about the star image centre.
Iterative ellipse fitting generates the accurate minor--axis FWHM, axis
ratio and orientation.  The pipeline logs the geometric mean of the
major and minor--axis FWHM in arcseconds.

Single--component models like the Gaussian or
even the twin--parameter S\'ersic function (S\'ersic 1968) are poor
representations of the UKIRT point--spread function under excellent
seeing conditions (see the UKIRT web pages, 
http://www.jach.hawaii.edu/JACpublic/UKIRT/telescope/telescope.html). 
Even in poor conditions the
Gaussian overestimates the width of the spiky inner core.
Nevertheless, for monitoring performance over a period, 
the Gaussian FWHM offers a
simple, easily interpreted statistic.

The results are rich in statistical information and are analyzed by a 
code produced by the authors. In order to avoid extended objects, we only 
employ observations of UKIRT 
faint standards. Finally, the code corrects the measurements to their 
equivalents at 1.0 airmasses and the K band, using standard dependences of 
the image size on airmass (i.e. seeing at zenith $\propto$ 
airmass$^{-3/5}$) and wavelength (i.e. seeing at 2.2$\mu$m $\propto$ 
$\left(\frac{\lambda}{2.2}\right)^{-1/5}$ $\times$ seeing at $\lambda$, where
$\lambda$ is measured in $\mu$m; see Sarazin \& Roddier 1990).

\begin{table}[h]
\caption{Median delivered image diameter as a function of the time of night in
2001. The standard error on each measurement of the median seeing is typically
$\pm$0.006.} 
\label{tab:dsee}
\begin{center}       
\begin{tabular}{|c|c|}
\hline
\rule[-1ex]{0pt}{3.5ex}  Hawaiian Standard  		& Median image \\
\rule[-1ex]{0pt}{3.5ex}  Time (HST) range          & diameter	\\
\rule[-1ex]{0pt}{3.5ex}  (hours)        & (arcsec)       \\
\hline
\rule[-1ex]{0pt}{3.5ex}  18--20	& 0.79	\\
\rule[-1ex]{0pt}{3.5ex}  20--22 & 0.65	\\
\rule[-1ex]{0pt}{3.5ex}  22--00	& 0.62	\\
\rule[-1ex]{0pt}{3.5ex}  00--02	& 0.59	\\
\rule[-1ex]{0pt}{3.5ex}  02--04	& 0.53	\\
\rule[-1ex]{0pt}{3.5ex}  04--06	& 0.54	\\
\rule[-1ex]{0pt}{3.5ex}  06--08	& 0.69	\\
\hline 
\end{tabular}
\end{center}
\end{table}

\begin{figure}
\includegraphics[height=20cm]{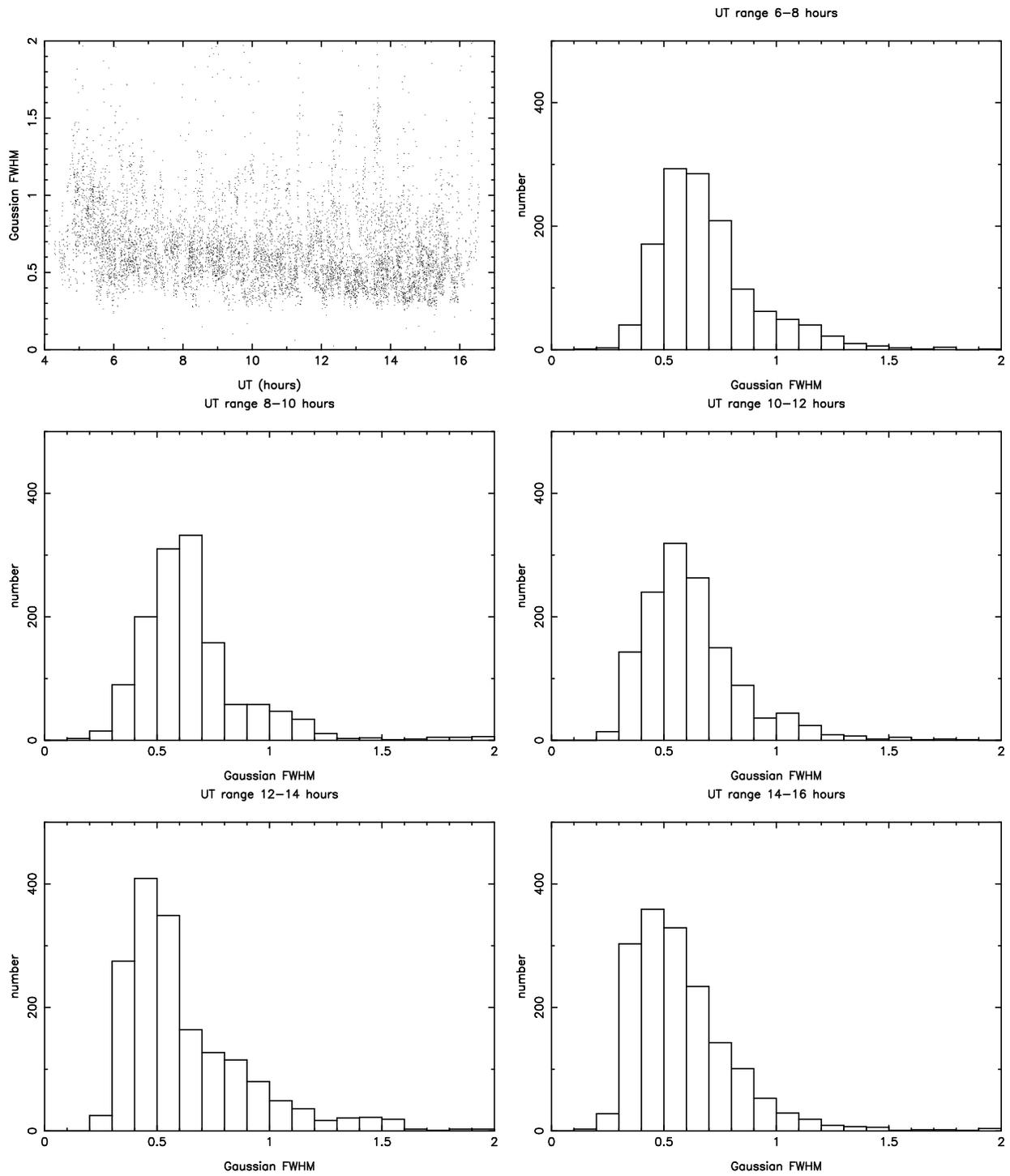}
\caption{Delivered image diameter versus time of the night, point by point 
(top left) and in histogram form.}
\end{figure}

UT coverage is divided into 5 main bins, 
each covering 2 hours from 6 to 16 hours 
UT. Image diameter 
is distributed into 20 bins, between 0$^{\prime\prime}$ and 
2$^{\prime\prime}$. The resulting 
histograms are shown in Figure 1, along with a plot of image diameter 
versus UT 
through the night. Table 1 also shows the median image quality 
for each UT bin and 
two extra UT bins at the beginning and end of the night. The additional bins, 
at HST 18--20 hours and HST 06--08 hours, both represent significantly smaller 
populations than the other UT bins and so their histograms are not shown in 
Figure 1.

These data show clearly the variation of image quality through the night. The 
histogram for the UT range 6--8 hours (HST 20--22 hours) 
in Figure 1 shows that by this time, the 
delivered image diameter 
has essentially stabilized, helped by the dome ventilation system. As 
the night progresses, the image quality 
gradually improves further. The best image quality 
is most often observed after 2am HST (12am UT). The histogram representing the 
UT range 12--14 hours demonstrates this clearly, with the counts quickly 
rising to 0.\hspace*{-1mm}$^{\prime\prime}$5, 
and then tailing off slowly. The median delivered image 
diameter for this time 
of the night drops to approximately 
0.\hspace*{-1mm}$^{\prime\prime}$53 (see Table 1).

Table 2 shows seasonal variations in the image quality. Each number in the 
table represents the percentage of each month for which the delivered FWHM 
was below a 
given value. August stands out as the best month, with delivered FWHM 
being better 
than 0.\hspace*{-1mm}$^{\prime\prime}$5 49\% of the time. 
July and September follow August as the next 
best months. At face value, November seems to be anomalous; but UFTI was 
rarely used during November, and then only towards the end of the month when 
conditions were very good.

\begin{table}
\caption{Shows how often the delivered image diameter 
was better than a particular value, e.g. in 
January the image diameter 
was better than 0.\hspace*{-1mm}$^{\prime\prime}$3 
0.48\% of the time, better than 0.\hspace*{-1mm}$^{\prime\prime}$4 3.8\% 
of the time, better than 0.\hspace*{-1mm}$^{\prime\prime}$5 
19.8\% of the time, etc. The last column lists 
the median seeing for that month. Data taken earlier than 6 
hrs UT are excluded.}
\begin{center}
\begin{tabular}{|l|l|l|l|l|l|l|}
\hline
Month	& \multicolumn{5}{c|}{Seeing (arcsec)}	   & Median Seeing\\
\hline
	& $\le$0.3  & $\le$0.4  & $\le$0.5  & $\le$0.6  & $\le$0.7  &		(arcsec)\\
\hline
Jan	& 0.48\% & 3.8\%  & 19.8\% & 41.8\% & 61.6\% & 0.63	\\
Feb	& 0.52 & 8.8  & 23.5 & 44.6 & 64.5 & 0.63	\\
Mar	& 0.80 & 1.2  & 6.8  & 20.1 & 45.6 & 0.71	\\
Apr	& 0.42 & 9.6  & 32.7 & 62.7 & 78.4 & 0.57	\\
May	& 1.09 & 9.0  & 34.0 & 59.2 & 74.0 & 0.58	\\
Jun	& 0.00 & 5.3  & 20.2 & 37.2 & 54.3 & 0.63	\\
Jul 	& 1.71 & 15.2 & 32.3 & 57.8 & 75.8 & 0.56	\\
Aug	& 2.84 & 27.3 & 49.0 & 64.1 & 79.7 & 0.50	\\
Sep	& 0.29 & 11.7 & 40.8 & 63.3 & 81.0 & 0.53	\\
Oct	& 1.09 & 9.4  & 20.7 & 29.7 & 36.6 & 0.82	\\
Nov	& 6.66 & 45.0 & 70.8 & 92.5 & 99.2 & 0.49	\\
Dec	& 0.00 & 2.2  & 11.2 & 27.7 & 46.0 & 0.74	\\
All	& 1.24 & 12.6 & 31.1 & 52.6 & 69.8 & 0.60	\\
\hline
\end{tabular}
\end{center}
\end{table}

The overall median image diameter 
for 2001, measured by Gaussian fitting as described 
above, was 0.\hspace*{-1mm}$^{\prime\prime}$6. 
While comparisons between these data and those obtained 
during the 1998 seeing campaign with IRCAM are probably invalidated by the use 
of different instrumentation, different methods of measuring the image quality 
(and of course the different epoch), there are some similarities, notably the 
distinct improvement in late summer.

\section{ATMOSPHERIC SEEING} 
\label{sec:asee}

Focus excursions measured by the autofocus system are a
measure of atmospheric seeing. Standard atmospheric theory predicts that the
seeing--induced image diameter is linearly related to the RMS 
atmospheric--induced focus fluctuations, $Z_{rms}$. In this mode, the fast
guider is acting as a form of Differential Image Motion Monitor (DIMM). 
$Z_{rms}$ should therefore
be a robust and invaluable measure of the atmospheric seeing (as opposed to
delivered image quality) at UKIRT.

The value of $Z_{rms}$ is measured every time the telescope is focussed.
During focussing, the Fast Guider re--images the telescope focal plane on the 
CCD using one of two lens systems mounted in a wheel. A single lens is 
employed for Normal Guide, Acquisition and Focus modes. For auto--focus a
second, similar lens is equipped with a 2$\times$2 
array of 4 f/100 Shack--Hartmann 
lenslets mounted immediately behind it. Instead of a single image these 
produce 4 images of the star, in a plane
somewhat nearer the lens wheel than the single image used for normal guiding. 

The longitudinal position of the image formed by the front lens on its own 
depends on the overall focus of the telescope. Consequently the diameter of the
converging pencil in any plane between the lens and its image
is a measure of the telescope focus. The four sub--images are formed from 
sub--pupils of the converging cone from the single lens, so the radial 
separation of the four images is also a measure of the overall telescope
focus setting. In auto--focus mode the four images are sensed by a 
24$\times$24 array of pixels, which are again binned up 3$\times$3, now into 
8$\times$8 superpixels forming four (rather than one) 4$\times$4 guiding 
arrays.

In autofocus mode the image positions are sensed at a rate of 60 Hz. 
These measurements are averaged over a specified interval, during which the 
RMS of the focus fluctuations, $Z_{rms}$, is also determined. Because the 
4$\times$4 sensor is non--linear at large excursions, such as might
accompany a change of instrument, when autofocus mode is initiated the 60--Hz 
focus corrections are averaged, and then applied, over consecutive periods of 
2, 4, 8, 16 and 32 seconds, first to facilitate convergence and then to allow 
seeing--induced fluctuations to be averaged out. The 32s averaging is then 
repeated indefinitely, until $Z_{rms}$ is seen to stabilise.

To convert $Z_{rms}$ to atmospheric
seeing the transformation shown in Equation 1 is used. In Equation 1, the
diffraction limit, 0.\hspace*{-1mm}$^{\prime\prime}$15 is added in quadrature 
with the contribution from $Z_{rms}$ to give the atmospheric seeing.

\begin{equation}
FWHM = \sqrt { 0.15^2 + ( 28.5 Z_{rms} ^ {6/5} ) ^2 }
\end{equation}

\begin{figure}[h]
\begin{center}
\includegraphics[height=10cm,angle=-90]{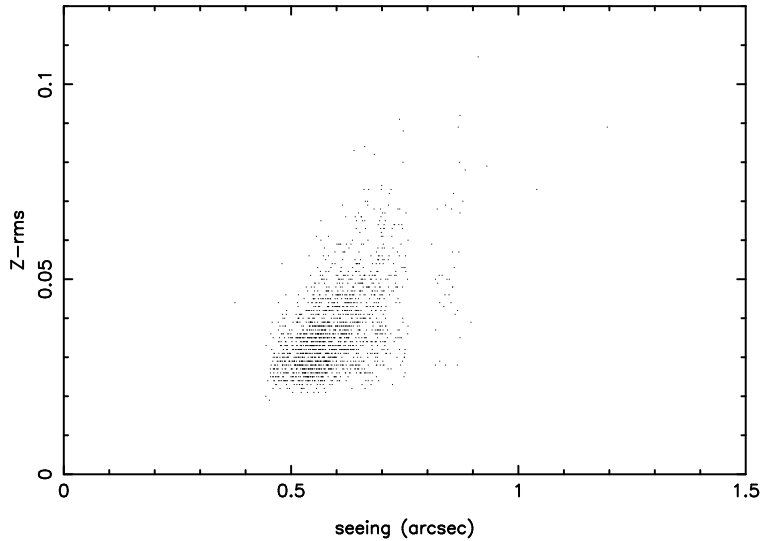}
\end{center}
\caption{$Z_{rms}$ versus atmospheric seeing for the nights of 1999 July 
29--30. A fit to these data results in the factor of 28.5 present in Equation 
1.}
\label{fig:massey}
\end{figure}

This equation is partly empirical, as the factor of 28.5 was measured by
directly comparing delivered image diameter with $Z_{rms}$ over 2 nights from
1999 July 29--30, when a single bright star was imaged in the {\em K} band
while simultaneously guiding in autofocus mode. The data from these two nights 
can be seen in Figure \ref{fig:massey}.

Using this transformation we have performed a similar analysis to that 
outlined in Section \ref{sec:dsee}, in order to see how well $Z_{rms}$
relates to the actual delivered seeing. In this analysis we have selected
values of $Z_{rms}$ from 2001, in order to remain consistent with the 
analysis of the delivered UFTI images. The $Z_{rms}$ logfile contains no
airmass information, and so the computed atmospheric seeing was not corrected
to zenith, although this is a small effect for all but extreme airmasses.

\begin{table}[h]
\caption{Median atmospheric seeing as a function of the time of night.} 
\label{tab:asee}
\begin{center}       
\begin{tabular}{|c|c|} 
\hline
\rule[-1ex]{0pt}{3.5ex}  HST range & Median seeing  \\
\rule[-1ex]{0pt}{3.5ex}  (hours)  & (arcsec)       \\
\hline
\rule[-1ex]{0pt}{3.5ex}  18--20	& 0.93$\pm$0.03	\\
\rule[-1ex]{0pt}{3.5ex}  20--22 & 0.82$\pm$0.02	\\
\rule[-1ex]{0pt}{3.5ex}  22--00	& 0.76$\pm$0.02	\\
\rule[-1ex]{0pt}{3.5ex}  00--02	& 0.58$\pm$0.02	\\
\rule[-1ex]{0pt}{3.5ex}  02--04	& 0.76$\pm$0.03	\\
\rule[-1ex]{0pt}{3.5ex}  04--06	& 0.78$\pm$0.04	\\
\hline 
\end{tabular}
\end{center}
\end{table}

\begin{figure}
\includegraphics[height=20cm]{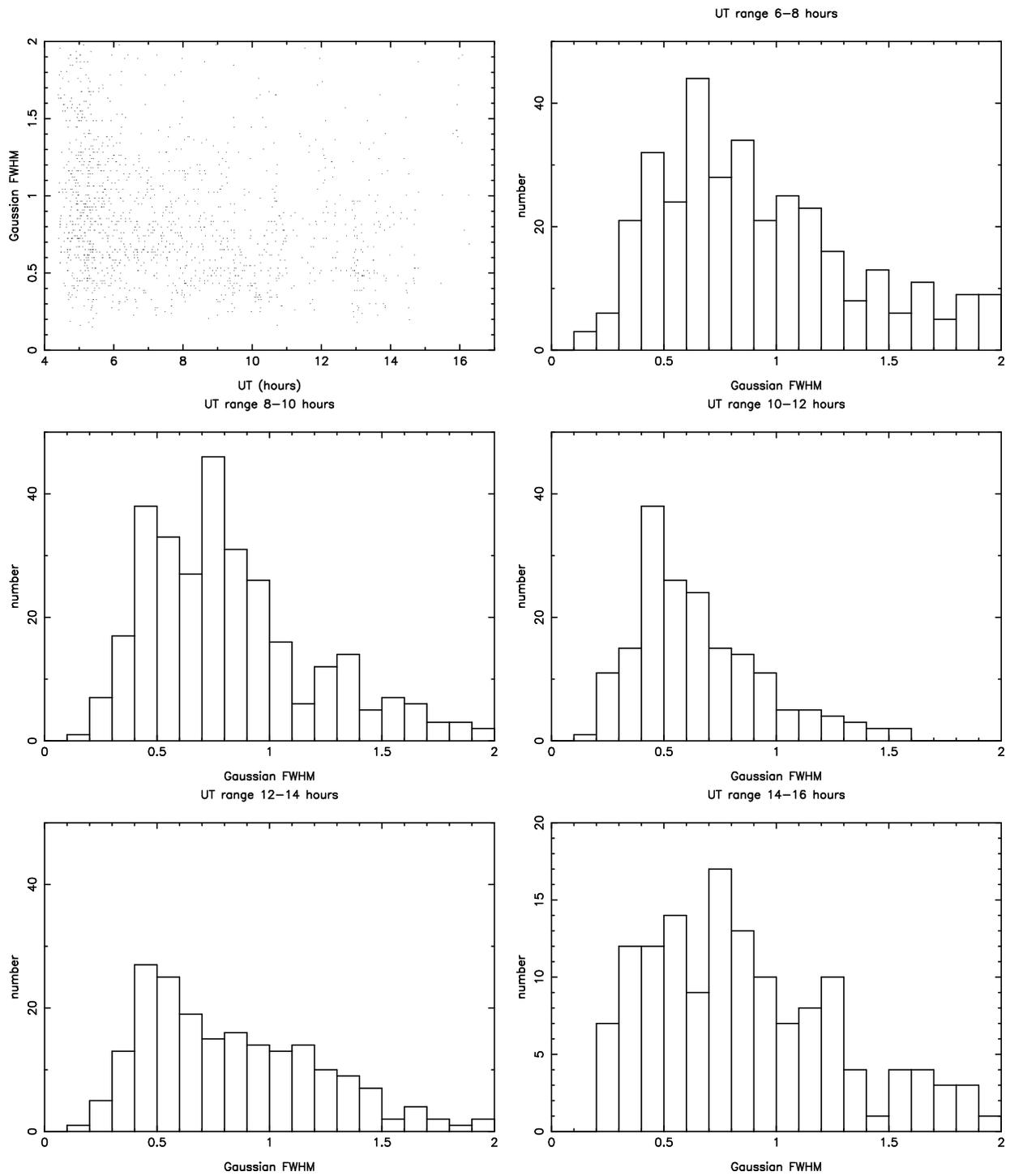}
\caption{{\em K} band seeing FWHM inferred from $Z_{rms}$ versus time of the 
night, point by point (top left) and in histogram form.}
\end{figure}

Once again, UT coverage is divided into 5 bins. 
The resulting histograms are shown in Figure 3, along
with a plot of atmospheric seeing versus UT through the night. Table 3 also 
shows the median seeing for each UT bin and one extra UT bin at the 
beginning of the night.

These data show how atmospheric seeing (as inferred from $Z_{rms}$)
varies through the night, although this
is clearly different from the delivered image quality. 
At all UT, the atmospheric
seeing values have a significantly larger spread than the delivered FWHM.
There is some evidence that the histograms are double peaked, except in the UT
range from 10 -- 12 hours. Even the median atmospheric seeing shown in Table 
3 does not show the same trend as the delivered seeing. In this case, the 
median does not slowly decrease through the night while the dome and telescope
structure are cooling. Instead a minimum is reached at 10 -- 12 hours, 
but then there is a sharp increase.
The overall inferred atmospheric seeing for 2001 was approximately 
0.\hspace*{-1mm}$^{\prime\prime}$8, 
significantly larger than the delivered image FWHM of 
0.\hspace*{-1mm}$^{\prime\prime}$6.



\section{DISCUSSION}
\label{sec:discuss}

As discussed in Section \ref{sec:asee} there is a theoretical link between
$Z_{rms}$ and seeing. Therefore, the delivered image quality should be linked
to the value of $Z_{rms}$ in some way as well. Measuring a diameter of a star
in a delivered image is not an estimate of the seeing.
Faults in the optical system of both the telescope and instrumentation will
degrade the best possible image by some degree. One would therefore expect
the FWHM of delivered images to be worse than the FWHM estimated from 
$Z_{rms}$. We would therefore, expect Figures 1 and 2 to look similar, but
with Figure 1 having a {\em higher} 
median value for each histogram than Figure 2.
This does not seem to be the case. The FWHM inferred from $Z_{rms}$ has both
a larger median and scatter than that measured in the delivered images.

When plotting the FWHM in the delivered images against $Z_{rms}$,
reasonable agreement is found with Equation 1. In Figure 5, the mean 
$Z_{rms}$ has been plotted versus 11 seeing bins 
(0.\hspace*{-1mm}$^{\prime\prime}$3--0.\hspace*{-1mm}$^{\prime\prime}$4, 
0.\hspace*{-1mm}$^{\prime\prime}$4--0.\hspace*{-1mm}$^{\prime\prime}$5 etc). 
This shows that, although
there is a large scatter in the value of the inferred seeing FWHM from 
$Z_{rms}$ compared to the delivered
image FWHM (see Figures 1 and 3), the mean value of $Z_{rms}$ in a large 
dataset tends towards value predicted by Equation 1. This suggests that 
the calibration used to predict Equation 1 is approximately correct. However,
the large scatter in $Z_{rms}$ is still a problem, especially for flexible
scheduling where the value of $Z_{rms}$ is often the quickest and simplest
way of flexing against the seeing. The cause of the larger median discussed
above is the long tails seen in the distributions of inferred atmospheric 
seeing in Figure 3.

\begin{figure}[h]
\begin{center}
\includegraphics[height=10cm,angle=-90]{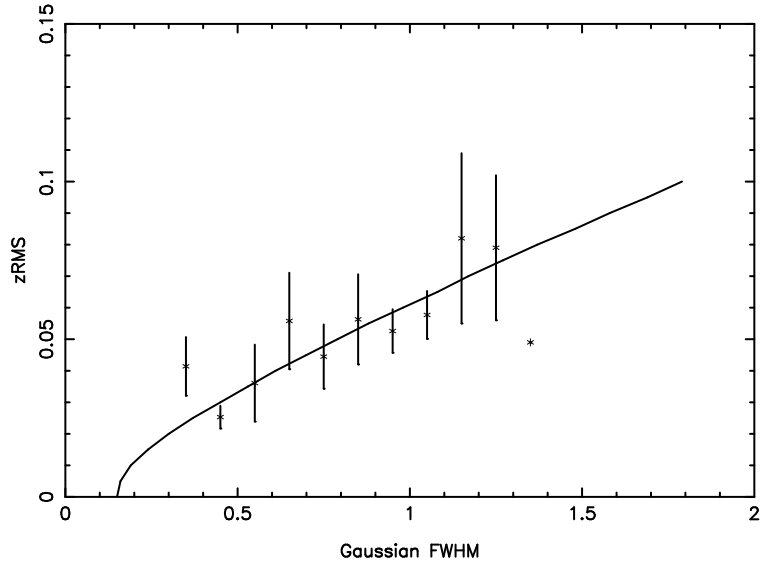}
\end{center}
\caption{$Z_{rms}$ versus FWHM in delivered images. The line represents
the theoretical link between $Z_{rms}$ and atmospheric seeing described by
Equation 1. The error bars are one standard error.}
\end{figure}

The intra--night variations for both the delivered 
FWHM and the $Z_{rms}$ inferred
FWHM have minima at some time after midnight. Table 1 shows that the delivered 
FWHM has a minimum value of 0.\hspace*{-1mm}$^{\prime\prime}$53 between 12--14 
hours UT,
rising to 
0.\hspace*{-1mm}$^{\prime\prime}$54 between 14--16 hours (although this rise
is not statistically significant). The minimum for 
atmospheric seeing occurs slightly earlier in the night, with a value
of 0.\hspace*{-1mm}$^{\prime\prime}$58 between 10--12 hours UT, and then a 
sharp rise to
0.\hspace*{-1mm}$^{\prime\prime}$76 between 12--14 hours.

\begin{figure}[ht]
\begin{center}
\includegraphics[height=10cm,angle=-90]{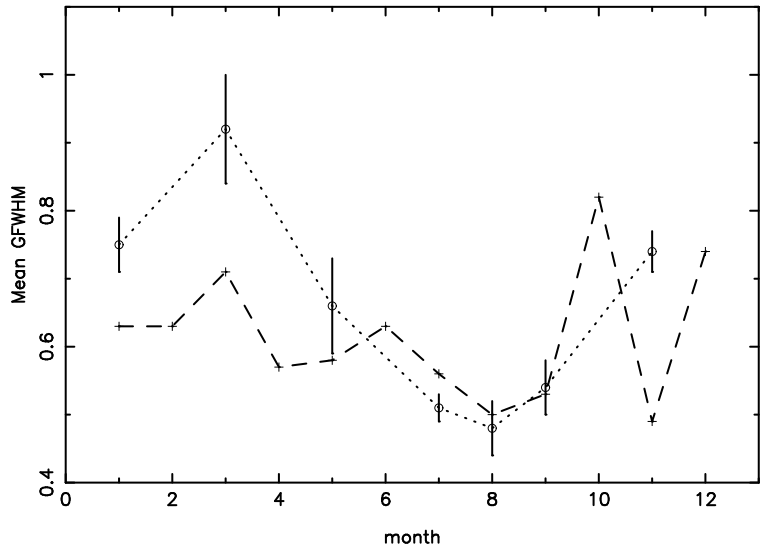}
\end{center}
\caption{Median delivered image quality versus month of the year (1=January, 2=February, 3=March etc). The dotted line represents 2000, with circles representing the data for 2000. The dashed line represents 2001, with crosses representing the data for 2001. The error bars on the data from 2001 are very small due to the large amount of data compared with 2000.}
\label{f:season}
\end{figure}

\begin{figure}[h]
\begin{center}
\includegraphics[height=10cm,angle=-90]{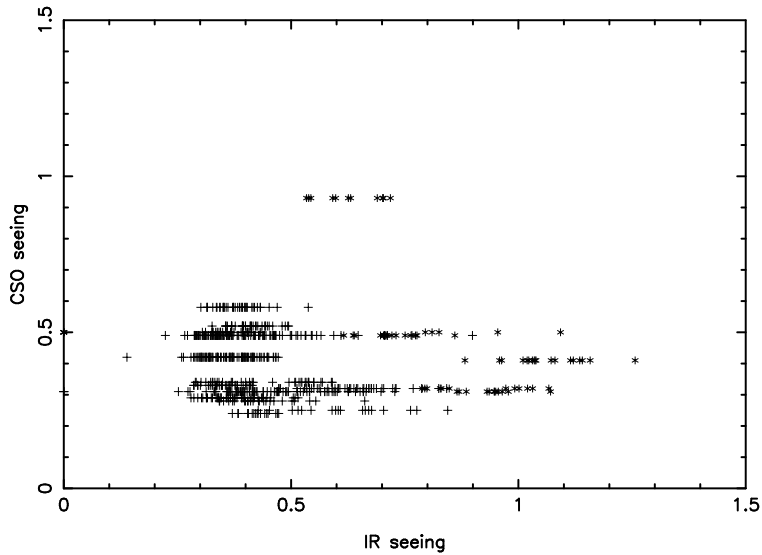}
\end{center}
\label{fig:irsee_vs_csosee}
\caption{CSO submillimetre seeing versus near--infrared K--band seeing. The
stars represent data taken on 2001 March 4th, which was a particularly bad
night for infrared seeing. The crosses represent data taken on 2001 August 
14th, which was a particularly good night for infrared seeing. In this 
plot, there are several measurements of IR seeing for every measurement
of CSO seeing.}
\end{figure}

Figure \ref{f:season} shows a plot of median delivered image FWHM 
versus the month of the
year for both 2000 and 2001. The statistics in Table 2, suggested that November
2001 was a very good month, but there were only a few nights available for 
observing in November 2001, due to bad weather conditions. It seems that the
nights that were not affected by the weather were anomalously good for that
particular time of year. The best seeing
for both years falls around the same time of year, i.e. from July to September.
This is consistent with the 
0.\hspace*{-1mm}$^{\prime\prime}$17 seeing images achieved
at UKIRT in September 1998 with IRCAM.
In 2000 a gradual increase is seen between the best months and November, 
whereas a discontinuous variation is seen between September and November
in 2001. This suggests that, as previously thought, November 2001 was 
anomalously good. There is anecdotal evidence to suggest that
very good seeing can 
occur directly after a storm, and this may be what we are seeing in November
2001.

Finally we have investigated the relationship 
between CSO submillimetre seeing and
near--infrared K--band seeing. We chose two nights on which to compare these
quantities, both of which were nights devoted to observations of UKIRT
faint standards. The first night, 2001 March 4th, was particularly bad for 
IR seeing. The second night, 2001 August 14th, was particularly good for IR 
seeing. The measurements
of IR seeing and CSO seeing for these nights are plotted in Figure 
\ref{fig:irsee_vs_csosee}, which shows no correlation between these two 
quantities. This is not surprising since submillimetre seeing depends mainly on
water vapour, whereas near--infrared seeing depends on atmospheric conditions
such as wind shear.

\section{CONCLUSIONS}

The main conclusions of this work are summarized as follows:

\begin{itemize}

\item The best seeing occurs in the second half of the night, generally after
2am HST.

\item The best seeing occurs in the summer, between the months of July and
September. This seems to be consistent with both 2000 and 2001, and also 1997,
when seeing of 0.\hspace*{-1mm}$^{\prime\prime}$17 was achieved in September.

\item The relationship between $Z_{rms}$ and delivered image diameter is 
uncertain. When mean $Z_{rms}$ is plotted versus seeing bins, the relationship
between the two fits the empirical model quite well. Therefore,
for a large dataset, the mean value of $Z_{rms}$ agrees well with the inferred
seeing, but in an individual measurement, the inferred seeing can disagree
with the measured image FWHM by up to $\sim$40\%, due to the large scatter in
$Z_{rms}$ with respect to the delivered image diameter. 
This impacts on how one uses
$Z_{rms}$ to predict the seeing with a flexibly scheduled telescope.

\item There is no correlation between CSO submillimetre seeing and
near--infrared seeing.

\end{itemize}

\acknowledgements
The United Kingdom Infrared Telescope is operated by the Joint Astronomy 
Centre, and Starlink is managed by the Council for the Central Laboratory for
the Research Councils, both on behalf of the U.K. Particle Physics and 
Astronomy Research Council.



\begin{thebibliography}{99}

\bibitem{Chrysostomou98}
Chrysostomou, A.C., Rees, N.P., Hawarden, T.G., Cavedoni, C.P., Pettie, D.G., Bennett, R.J., Atad--Ettedgui, E., Humphries, C.M., \& Mack, B., 1998, {\em Active optics at UKIRT.} Proc SPIE, 3352, 446

\bibitem{Currie01}
Currie, M.J., 2001, Starlink User Note 232\\
{\tt http://www.starlink.rl.ac.uk/star/docs/sun232.htx/sun232.html}

\bibitem{Currie00}
Currie, M.J., \& Berry, D.S., 2000, Starlink User Note 95\\
{\tt http://www.starlink.rl.ac.uk/star/docs/sun95.htx/sun95.html}

\bibitem{Economou99}
Economou, F., Bridger, A., Wright, G.S., Jenness, T., Currie, M.J., \& Adamson, A.J., 1999, in ASP Conf. Ser., Vol. 172, Astronomical Data Analysis Software and Systems VIII, eds. D.M. Mehringer, R.L. Plante, \& D.A. Roberts (San Francisco: ASP), 11 

\bibitem{Hawarden94}
Hawarden, T.G., Cavedoni, C.P., Rees, N.P., Chuter, T.C., Pettie, D.G., Humphries, C.M., Bennett, R.J., Atad, E., Harris, J.W., Mack, B., Pitz, E., Glindeman, A., \& Rohloff, R.--R., 1994, {\em The UKIRT Upgrades Programme: Preparing for the 21st Century.} Proc SPIE, 2199, 494

\bibitem{Hawarden96}
Hawarden, T.G., Cavedoni, C.P., Chuter, T.C., Look, I.A., Rees, N.P., Pettie, D.G., Bennett, R.J., Atad, E., Harris, J.W., Humphries, C.M., Mack, B., Pitz, E., Glindeman, A., Hippler, S., Rohloff, R.--R., \& Wagner, K., 1996, {\em Progress of the UKIRT Upgrades Programme.} Proc SPIE, 2871, 256

\bibitem{Hawarden98}
Hawarden, T.G., Rees, N.P., Cavedoni, C.P., Chuter, T.C., Chrysostomou, A.C., Pettie, D.G., Bennett, R.J., Atad, E., Harris, J.W., Mack, B., Pitz, E., Glindeman, A., Hippler, S., Rohloff, R.--R., \& Wagner, K., 1998, {\em The Upgraded UKIRT.} Proc SPIE, 3352, 52

\bibitem{Hawarden99}
Hawarden, T.G., Rees, N.P., Chuter, T.C., Chrysostomou, A.C., Cavedoni, C.P., Rohloff, R.--R., Pitz, E., Pettie, D.G., Bennett, R.J., \& Atad--Ettedgui, E., 1999, {\em Post--Upgrade Performance of the 3.8m UK Infrared Telescope (UKIRT).} Proc SPIE, 3785, 82

\bibitem{Hawarden00}
Hawarden, T.G., Adamson, A.J., Chuter, T.C., Rees, N.P., Massey, R.J., Cavedoni, C.P., \& Atad--Ettedgui, E., 2000, {\em Thermal Performance and Facility Seeing at the Upgraded 3.8m UK Infrared Telescope (UKIRT).} Proc SPIE, 4004, 104

\bibitem{sarazin90}\
Sarazin, M., \& Roddier, F., 1990, {\em The ESO differential image motion monitor.} A\&A, 227, 294

\bibitem{sersic68}
S\'ersic, J.--L., 1968, Atlas de galaxias australes (Observatorio Astronomica, Cordoba)

\end{thebibliography}
\end{document}